\documentclass[referee]{aa}
\usepackage{graphicx}
\usepackage{txfonts}
\usepackage{cases}

\begin{document}
   \title{Reciprocatory magnetic reconnection in a coronal bright point}
   
      \author{Q. M. Zhang\inst{1,3},
         P. F. Chen\inst{2,3},
         M. D. Ding\inst{2,3},
         \and
         H. S. Ji\inst{1}
          }

   \institute{Key Laboratory for Dark Matter and Space Science, Purple
              Mountain Observatory, CAS, Nanjing 210008, China\\ 
              \email{zhangqm@pmo.ac.cn}
        \and
              School of Astronomy and Space Science, Nanjing University,
              Nanjing 210093, China
         \and
             Key Lab of Modern Astronomy and Astrophysics, Ministry of
              Education, China\\
             }

   \date{Received; accepted}
    \titlerunning{Magnetic reconnection in a coronal bright point}
    \authorrunning{Zhang et al.}

  \abstract
   {Coronal bright points (CBPs) are small-scale and long-duration brightenings 
   in the lower solar corona. They are often explained in terms of magnetic 
   reconnection.}
   {We aim to study the sub-structures of a CBP and clarify the relationship among the 
   brightenings of different patches inside the CBP.}
   {The event was observed by the X-ray Telescope (XRT) aboard the Hinode spacecraft 
   on 2009 August 22$-$23.}
   {The CBP showed repetitive brightenings (or CBP flashes). During each of the two  
   successive CBP flashes, i.e., weak and strong flashes which are separated by $\sim$2 hr, 
   the XRT images revealed that the CBP was composed of two chambers, i.e., patches A and B. 
   During the weak flash, patch A brightened first, and patch B brightened $\sim$2 min later. 
   During the transition, the right leg of a large-scale coronal loop drifted from the right side of 
   the CBP to the left side. During the strong flash, patch B brightened first, and patch A 
   brightened $\sim$2 min later. During the transition, the right leg of the large-scale coronal 
   loop drifted from the left side of the CBP to the right side. In each flash, the rapid change 
   of the connectivity of the large-scale coronal loop is strongly suggestive of the interchange 
   reconnection.}
   {For the first time we found reciprocatory reconnection in the CBP, i.e., reconnected 
   loops in the outflow region of the first reconnection process serve as the inflow of the second 
   reconnection process.}


   \maketitle

\section{Introduction}

Coronal bright points (CBPs) are ubiquitous small-scale (10$\arcsec$$-$40$\arcsec$) 
brightenings in the lower corona (Krieger et al. \cite{kri71}; Zhang et al. 
\cite{zhang01}; Kariyappa \& Varghese \cite{kari08}; Huang et al. \cite{hua12}; 
Kwon et al. \cite{kwon12}), which may or may not have H$\alpha$ counterparts 
(Zhang et al. \cite{zhang12}). The point-like or loop-like bright features are widely 
distributed in the quiet regions as well as coronal holes (Habbal et al. \cite{harb90}; 
Kotoku et al. \cite{kot07}; Madjarska \& Wiegelmann \cite{mad09}; Zhang \& Ji 
\cite{zqm13}). The temperature and density of CBPs are typically 1$-$3 MK and 
10$^9$$-$10$^{10}$ cm$^{-3}$ (Ugarte-Urra et al. \cite{uga05}; Kariyappa et al. 
\cite{kari11}). The lifetimes of CBPs (2$-$48 hr) are proportional to the magnetic 
flux of the corresponding bipoles in the photosphere (Golub et al. \cite{golub77}) 
as well as the EUV intensities (Chandrashekhar et al. \cite{chan13}). At the boundaries 
of active regions and coronal holes, CBPs are often associated with fast jets when 
newly emerging flux encounters pre-existing magnetic field lines with opposite polarity
and interchange magnetic reconnection occurs, which plays an important role in the 
dynamic evolution of coronal holes and slow solar wind (Culhane et al. \cite{cul07}; 
Filippov et al. \cite{fili09}; Subramanian et al. \cite{sub10}; Madjarska et al. \cite{mad12}; 
Lee et al. \cite{lee13}; Archontis \& Hood \cite{arc13}; Moore et al. \cite{moor10, moor13}; 
Moreno-Insertis \& Galsgaard \cite{more13}; Zhang \& Ji \cite{zqm14a,zqm14b}). It has been 
proposed that magnetic reconnection is responsible for the heating of CBPs (Priest et al. 
\cite{pri94}; Parnell et al. \cite{par94}; van Driel-Gesztelyi et al. \cite{van96}; Longcope 
\cite{lon98}; Santos et al. \cite{sant08}). However, the corresponding magnetic 
configuration, which distinguishes CBPs from ordinary flares and microflares, has not 
been investigated with sufficient efforts. Zhang et al. (\cite{zqm12}) studied the soft X-ray 
(SXR) evolution and magnetic topology of two CBPs. Both of them were associated with 
``embedded bipolar magnetic field'', implying a magnetic null point in the corona, which 
was further confirmed by potential-field extrapolation. Considering that the light curves 
of the two CBPs consist of repetitive flashes superposed over long-existing brightening
(Habbal \& Withbroe \cite{harb81}; Tian et al. \cite{tian08}), the authors proposed that the 
CBP evolution consists of two components: quasi-periodic impulsive flashes and gradual 
weak brightening, which correspond to null-point and quasi-separatrix layer (QSL) 
reconnections, respectively. The null-point reconnection produces jets associated with 
CBP flashes, whereas QSL reconnection is much gentler (Mandrini et al. \cite{man96}; 
Aulanier et al. \cite{aul07}; Pontin et al. \cite{pont11}; Guo et al. \cite{guo13}; 
Schmieder et al. \cite{sch13}). Although not all CBPs have a null point and fan-spine
magnetic configuration, the two components of SXR brightening seem to exist and two 
types of magnetic reconnection, i.e., anti-parallel and slip-running reconnection in QSLs, 
apply to most CBPs. 

However, the above model has not explained the formation of sub-structures
inside a CBP. High-resolution observations revealed that CBPs are often composed 
of a bundle of bright loops (Sheeley \& Golub \cite{shee79}; Dere \cite{dere08}; 
P{\'e}rez-Su{\'a}rez et al. \cite{per08}). 
It is still a puzzle how the sub-structures inside a CBP are related to each other. 
With this in mind, we studied a CBP observed by the X-ray Telescope (XRT; 
Golub et al. \cite{golub07}) aboard the Hinode spacecraft (Kosugi et al. 
\cite{kos07}). In Sect. \ref{s-data}, we describe the observation and data analysis. 
In Sect. \ref{s-result}, we show the results. The discussion and summary are 
presented in Sect. \ref{s-disc}.

\section{Observation and Data Analysis} \label{s-data}

\begin{figure}
\includegraphics[width=12cm,clip=]{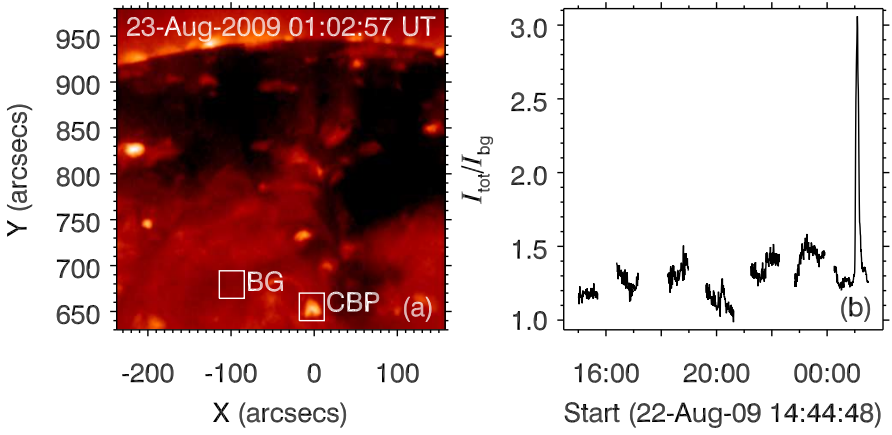}
\caption{{\bf a)} A partial-frame SXR image observed by XRT at 01:03 UT on
August 23. The two white boxes stand for the background region (BG) and the
CBP. {\bf b)} SXR light curve of the CBP. The light curve is calculated to be
the total flux of the CBP normalized by the total flux of BG. Note the data
gap in panel {\bf b)}.}
\label{fig1}
\end{figure}

XRT conducted partial-frame (384$\arcsec\times384\arcsec$) observations from 
2009 August 22 15:00 UT to August 23 01:30 UT, with a time cadence of 32 s
and a pixel size of 1\farcs03. The raw data are calibrated using {\it xrt\_prep.pro}, 
a routine in Solar Software. As shown by Fig. \ref{fig1}a, the XRT telescope was 
observing a small region centered at (-40$\arcsec$, 780$\arcsec$), which is
close to the north polar coronal hole. A CBP was located near (-3$\arcsec$,
655$\arcsec$) as indicated by the small box. Its light curve is plotted in
Fig. \ref{fig1}b. After checking the XRT movie, we find that the CBP appeared 
as a small loop during 15:00$-$22:16 UT on August 22. From 22:49 UT on
August 22 to 01:29 UT on August 23, the CBP experienced two flashes and 
presented interesting structures. We focus on this time span and divide it into
weak flash phase (from 22:49 UT on August 22 to 00:40 UT on August 23) and
strong flash phase (00:40$-$01:29 UT on August 23) during which the CBP 
induced a sympathetic CBP and gentle chromospheric evaporations within it 
(Zhang \& Ji \cite{zqm13}). Since alternative brightening of two patches was 
observed inside the CBP, it is more appropriate to call it a CBP complex, rather 
than a simple one. In this paper, we investigate the magnetic reconnection 
within the CBP itself. We tried to figure out the three-dimensional magnetic 
configuration of the CBP by performing potential-field extrapolation based on 
the line-of-sight magnetograms from the Big Bear Solar Observatory but failed 
owning to the extremely weak and noisy magnetic fields near the polar region.
We also examined 
the Transition Region And Coronal Explorer (TRACE; Handy et al. \cite{hand99}) 
data. Unfortunately, the 171 {\AA} images during that time are too poor with low 
time cadence ($>$2 min) and too much noise, so that they are not helpful for our 
research.

\section{Results} \label{s-result}

\subsection{Weak Flash Phase} \label{sub-w}

\begin{figure}
\includegraphics[width=10cm,clip=]{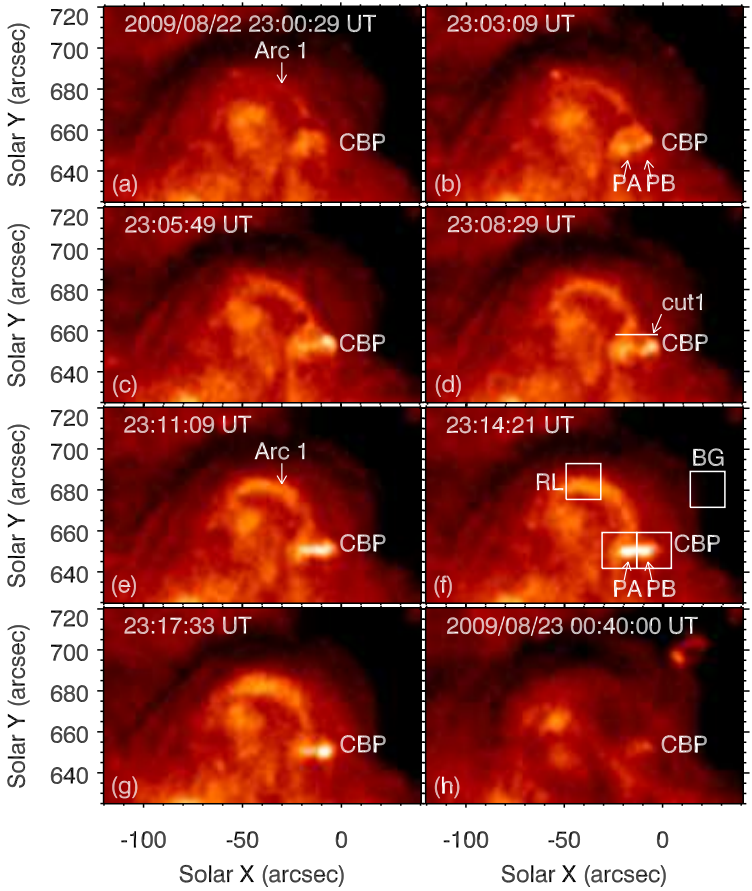}
\caption{{\bf a)$-$h)} Eight snapshots of the SXR images during the 
weak flash phase of the CBP. 
PA and PB stand for patch A and patch B of of the CBP. ``Arc 1''
signifies the magnetic loop connecting the CBP with the remote leg (RL). 
BG denotes the background region used for the calculation of light curves in
Fig.~\ref{fig3}. The white solid line denoted with ``cut1'' in panel {\bf d)} 
is used for studying the transverse drift of ``Arc 1''. An animation of this 
figure ({\it animation1.mov}) is available in the online journal.}
\label{fig2}
\end{figure}

Figure~\ref{fig2} displays the evolution of CBP complex during the weak flash
phase. Initially at 23:00:29 UT, it looked like a small diffuse loop with a
horizontal size of 18$\arcsec$ located to the left of (-10$\arcsec$, 
650$\arcsec$). A long arc-shaped bright loop, ``Arc 1'' hereafter, became
evident, linking the right side of CBP to northeast. Three minutes later at
23:03:09 UT, the small diffuse loop, which is labeled ``patch A (or PA)" of
the CBP became smaller and smaller, and another bright patch, which is labeled
``patch B (or PB)" of the CBP became visible to the right of patch A near 
(-8$\arcsec$, 655$\arcsec$). At the same time, the bright loop ``Arc 1''
suddenly changed its connectivity at the western leg from the right side of
the CBP to the left side. As seen from the attached movie, ``Arc 1'' 
detached from the solar surface during the transition and reconnected to
the upper loop of CBP, forming a kinked loop structure resembling the number
``7''. At 23:14:21 UT, patch A reached maximum in intensity and it was
apparently connected to patch B, while the remote or 
eastern leg of ``Arc 1'' also became the brightest. Afterwards, patch A of
CBP decayed rapidly, whereas patch B and the remote leg of ``Arc 1''
decayed much more slowly. At 00:40 UT on August 23, both patch A 
and ``Arc 1'' were invisible, and patch B was diffuse and faint. 

\begin{figure}
\includegraphics[width=12cm,clip=]{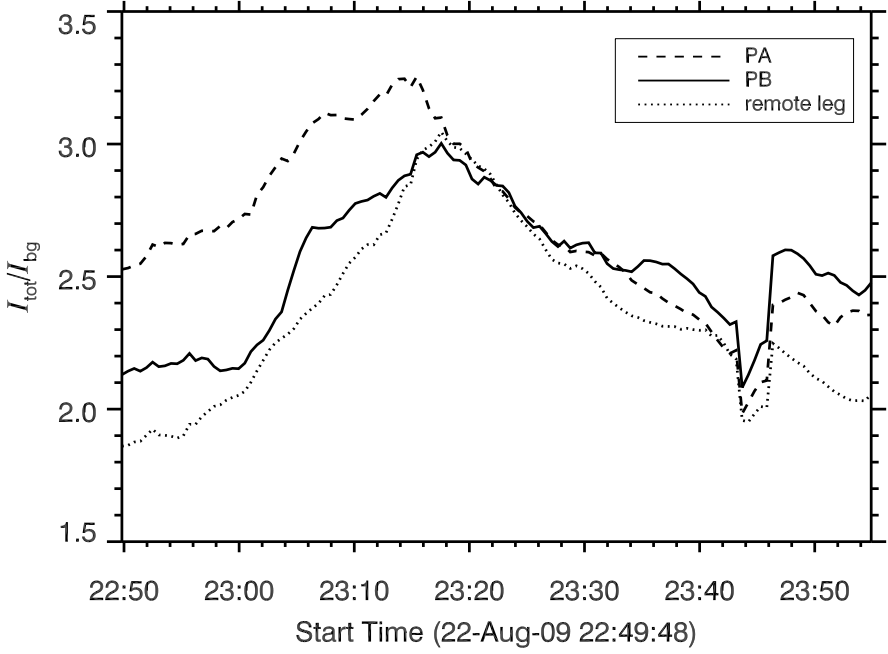}
\caption{SXR light curves of patch A (PA) ({\it dashed line}), 
patch B (PB) ({\it solid line}), and the remote leg of ``Arc 1'' 
({\it dotted line}) during the weak flash phase of the CBP.}
\label{fig3}
\end{figure}

In order to illustrate the temporal relationship among the brightness of the
two patches of CBP and the remote leg of large-scale loop ``Arc 1'', the
SXR light curves of the three parts are plotted in Fig.~\ref{fig3}. The
intensity for each part is the integral flux inside the corresponding boxes in
Fig.~\ref{fig2}, which is normalized by the total flux of a background region 
with the same area. As indicated by Fig.~\ref{fig3}, patch A started to 
brighten at $\sim$22:50 UT and reached its peak around 23:15 UT. The 
intensity of patch B increased from $\sim$23:00 UT and peaked around 23:17 
UT on August 22. The time delay between the peaks of the two patches was 
$\sim$2 min. However, the maximum intensity of the remote leg of ``Arc 1'' 
was nearly simultaneous with that of patch B, implying the delay between them 
is less than the time cadence of the XRT observation (i.e., $<$32 s).

\begin{figure}
\includegraphics[width=10cm,clip=]{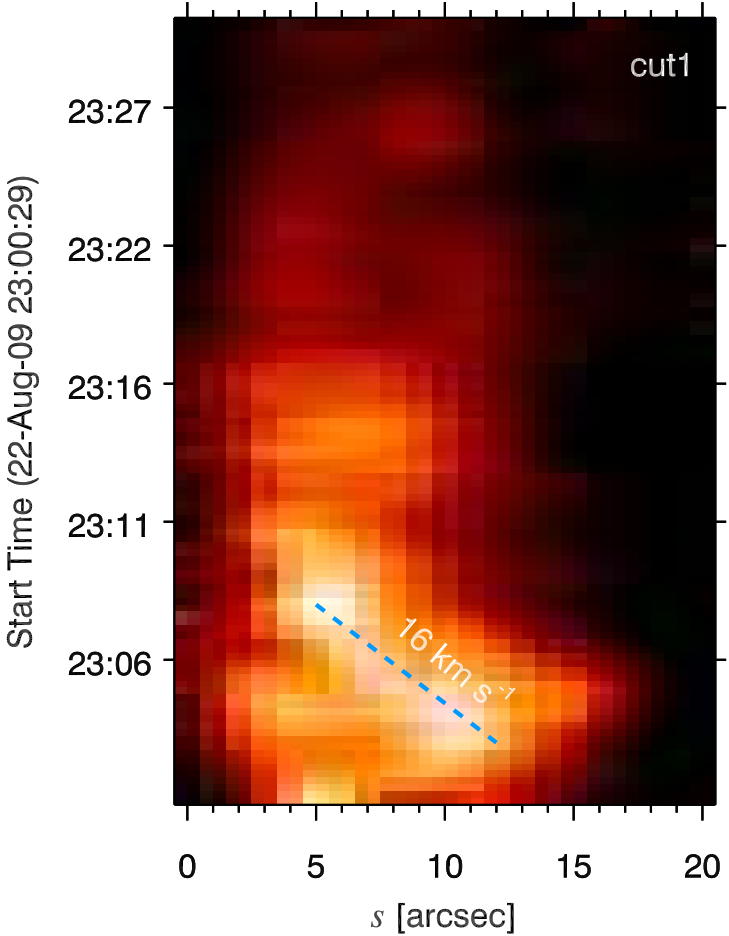}
\caption{Time-slice diagram of the SXR intensity along cut1 during the weak flash phase. The 
blue dashed line stands for the leftwards drift of ``Arc 1''. The slope of the blue 
line represents the apparent drift speed (16 km s$^{-1}$).}
\label{fig4}
\end{figure}

To better illustrate the transition of the connection of ``Arc 1'', we extracted 
a cut across ``Arc 1'', i.e., ``cut1" in Fig.~\ref{fig2}d. The time-slice diagram of cut1 is 
displayed in Fig.~\ref{fig4}. It shows that the western leg of ``Arc 1'' drifted leftwards
at an apparent speed of $\sim$16 km s$^{-1}$ during the weak flash phase, 
as indicated by the blue dashed line.

\subsection{Strong Flash Phase} \label{sub-S}

\begin{figure}
\includegraphics[width=10cm,clip=]{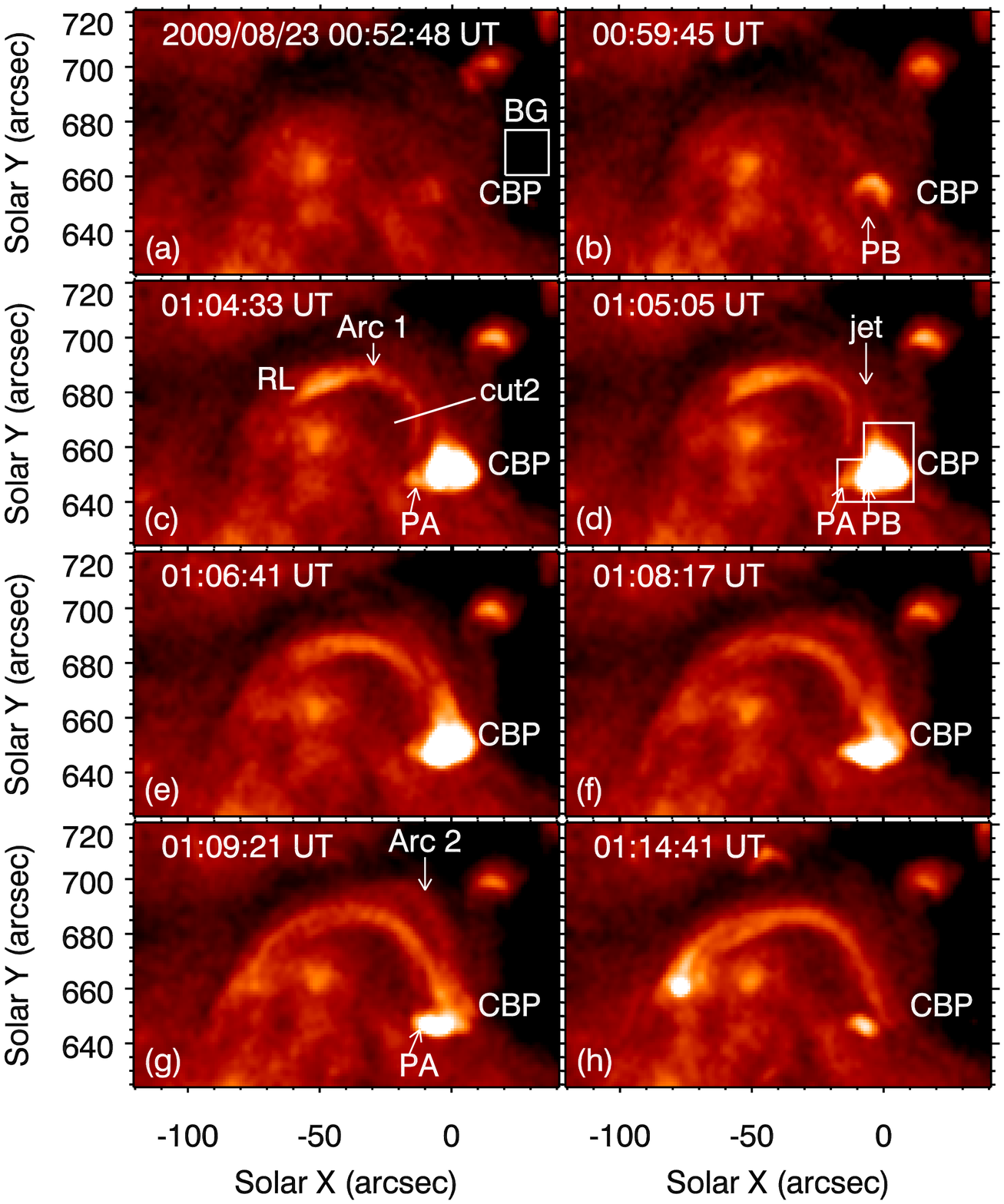}
\caption{{\bf a)$-$h)} Eight snapshots the SXR images during the strong 
flash phase of the CBP. PA, PB, ``Arc 1'', and RL have the same meanings 
as in Fig.~\ref{fig2}. ``Arc 2'' signifies the magnetic loop adjacent to ``Arc 1''. 
The white solid line denoted with ``cut2'' in panel {\bf c)} is used for studying 
the transverse drift of ``Arc 1''. An animation of this figure 
({\it animation2.mov}) is available in the online journal.}
\label{fig5}
\end{figure}

After 00:40 UT on August 23, patch B of the CBP was torpid for $\sim$12 min
before shining rapidly, with the brightness being stronger than ever. The
evolution of CBP in the strong flash phase is demonstrated in Fig.~\ref{fig5}. 
At 00:59:45 UT, only patch B was identifiable, which looked like a
small bright loop. Afterwards, its brightness increased drastically. When
patch B flashed, a tiny region to the left of patch B became visible, which is
indicated in Fig.~\ref{fig5}c. Since the tiny bright region was located at
the same site as patch A in the weak flash phase mentioned in Sect. \ref{sub-w},
we consider the tiny bright point as the recurrence of patch A of the CBP.
Meanwhile, a brightening propagates rapidly along an arc-shaped loop from 
the top left of patch B to the remote site. Based on the same connectivity 
and cospatiality, we consider the large bright loop as the recurrence of ``Arc
1''. At 01:05:05 UT, a jet-like structure emanated from the top-right of patch
B and propagated along another bright loop ``Arc 2'' that was slightly shifted
from ``Arc 1'' in Fig.~\ref{fig5}d. At 01:06:41 UT, ``Arc 1'' suddenly
changed its connectivity at the western leg from the left side of the CBP to the
right side. At 01:09:21 UT, ``Arc 1'' was clearly rooted to the right side of
patch B. At the same time, the left footpoint of patch B merged with patch A.
After that, patch B disappeared rapidly, while patch A and ``Arc 1'' decayed
slowly.

\begin{figure}
\includegraphics[width=12cm,clip=]{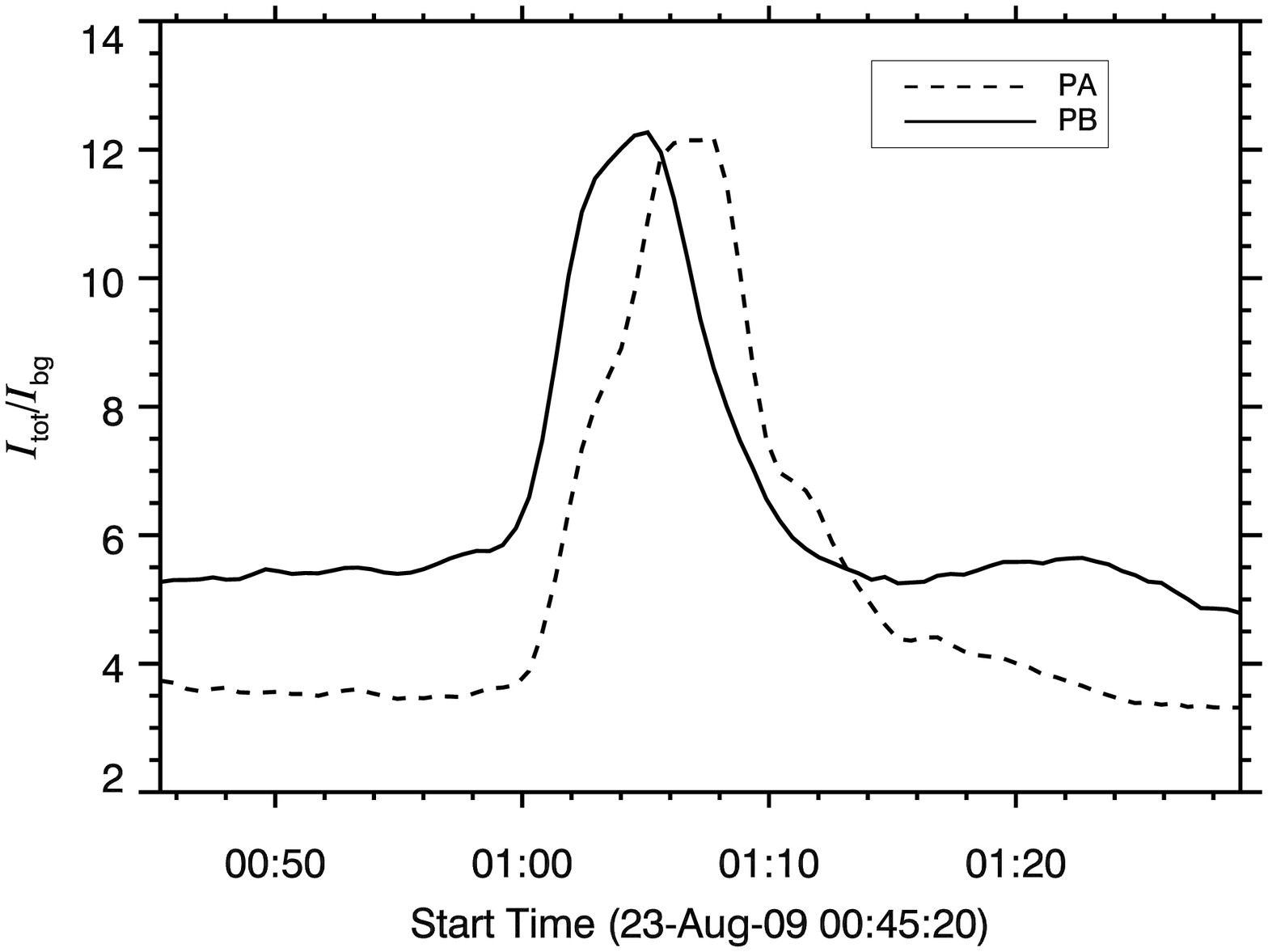}
\caption{The SXR light curves of patch A (PA) ({\it dashed line}) and 
patch B (PB) ({\it solid line}) during the strong flash phase of the CBP.
The intensity of PA is multiplied by 1.8 to get a better comparison with 
PB.}
\label{fig6}
\end{figure}

As in the weak flash phase, we investigate the temporal relationship 
between patches A and B. Noticing that it is difficult to distinguish the two 
patches when the total intensity of patch B peaked, we select two boxes in 
Fig.~\ref{fig5}d, covering the main parts of patches A and B, respectively. The
SXR flux inside the boxes are integrated separately and then normalized by the
total SXR flux in the selected background area BG. The light curves of the two
boxes representing patches A and B are plotted in Fig.~\ref{fig6}. It is shown
that patch B brightened first from $\sim$00:58 UT and reached its peak at
$\sim$01:05 UT on August 23. The intensity of patch A increased from 01:00 UT
and peaked around 01:07 UT on August 23. The time delay between the two 
patches, $\sim$2 min, was very much similar to that during the weak flash phase.

\begin{figure}
\includegraphics[width=10cm,clip=]{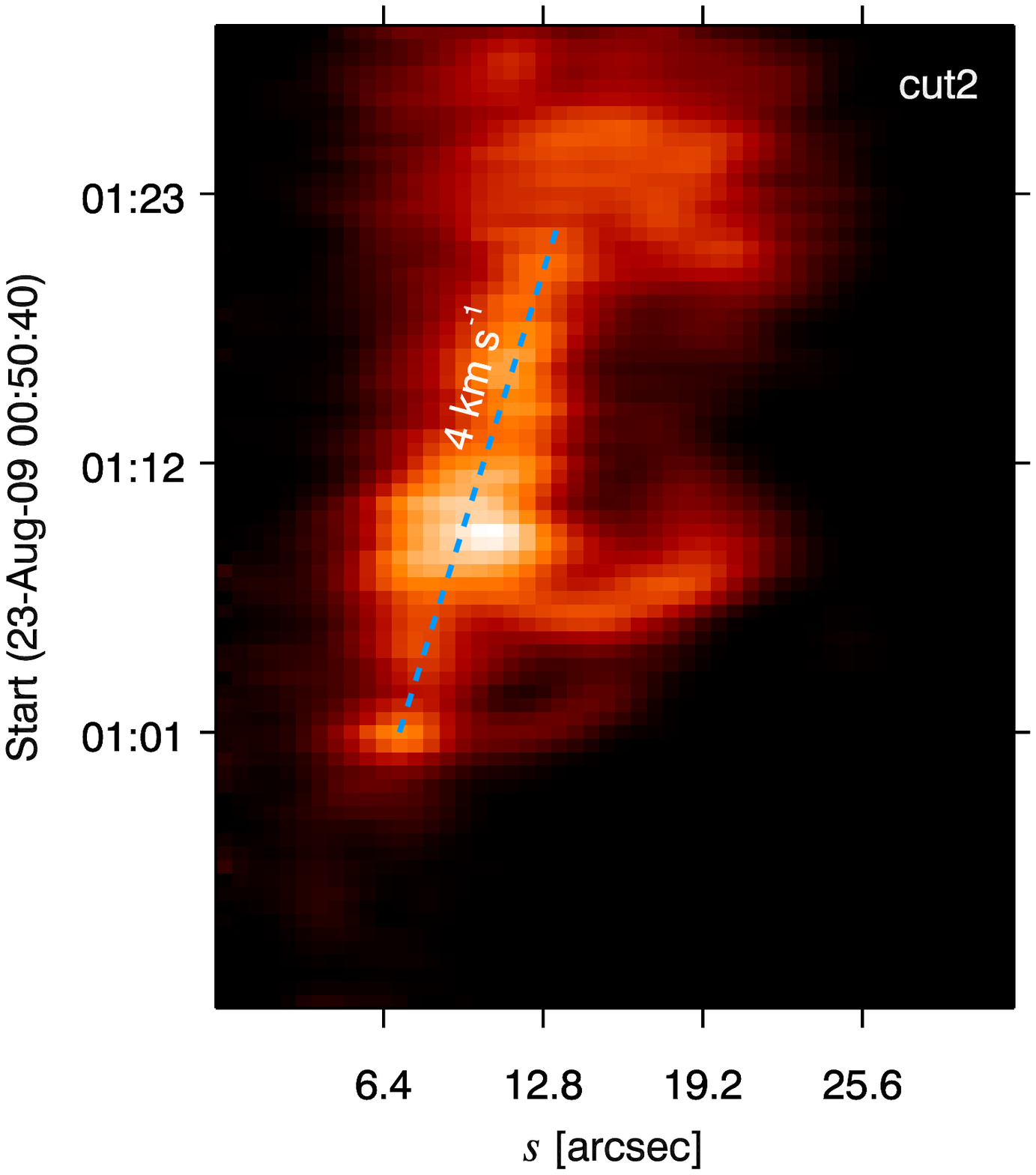}
\caption{Time-slice diagram of the SXR intensity along cut2 during the strong flash 
phase. The blue dashed line stands for the rightwards drift of ``Arc 1''. The slope 
of the blue line represents the apparent drift speed (4 km s$^{-1}$).}
\label{fig7}
\end{figure}

Like in Fig.~\ref{fig2}d, we extracted another cut across ``Arc 1'', i.e., cut2 in 
Fig.~\ref{fig5}c. The time-slice diagram of cut2 is displayed in Fig.~\ref{fig7}. It  
reveals that the western leg of ``Arc 1'' drifted rightwards at an apparent speed
of $\sim$4 km s$^{-1}$ during the strong flash phase.

\section{Discussion and Summary} \label{s-disc}

Magnetic reconnection has been proposed to explain CBPs for a long time via
different models where the differences between these models lie in the
magnetic configuration involved in the reconnection. Based on the light curves
and potential fields of two CBPs, Zhang et al. (\cite{zqm12}) proposed that the 
embedded bipolar magnetic configuration might be responsible for the typical 
CBPs whose brightening consists of a weak gradual component and a quasi-periodic
impulsive component, the latter was also called CBP flashes. According to the
model, QSL reconnection or component reconnection accounts for the gradual
component of brightening, whereas null-point reconnection accounts for the CBP
flashes in nearly the same way as compact flares (Heyvaerts et al. \cite{heyv77}) 
or coronal jets (Shibata et al. \cite{shib92}): A small bipolar magnetic loop (e.g., an 
emerging flux) reconnects with an overlying large loop or open fields. As mentioned 
by Moreno-Insertis et al. (\cite{more08}), this kind of reconnection is characterized by 
two chambers, a shrinking one containing the emerging magnetic loops and a growing 
one with loops produced by reconnection. They are located in the reconnection inflow
and outflow regions. The two-chamber structure is discernible as patches A and B in
Figs.~\ref{fig2} and \ref{fig5}. However, the role of each chamber in the two
phases might be completely different.

\begin{figure}
 \centerline{\hspace*{0.01\textwidth}
             \includegraphics[width=0.44\textwidth,clip=]{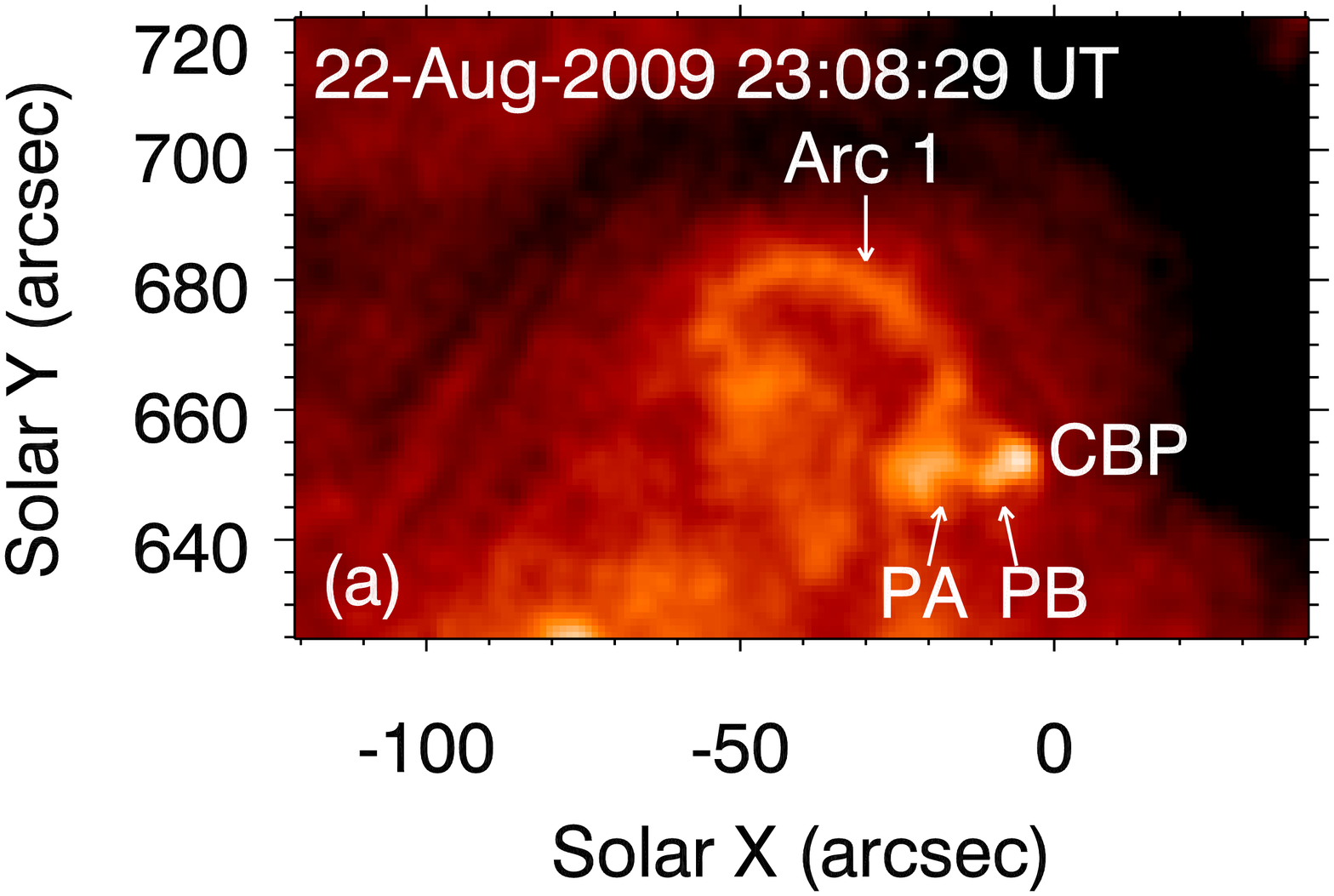}
             \hspace*{0.01\textwidth}
             \includegraphics[width=0.44\textwidth,clip=]{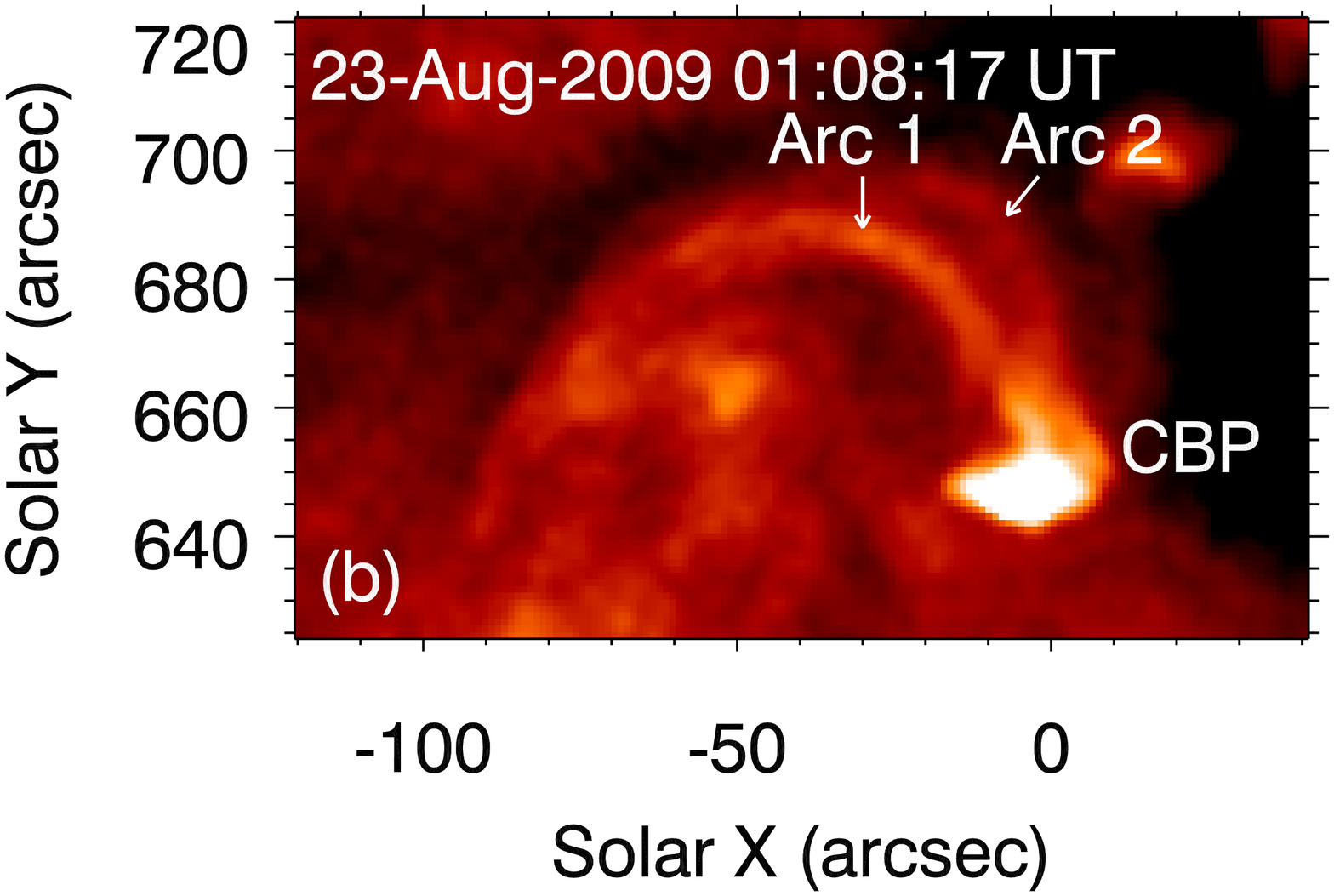}
            }

   \centerline{\hspace*{0.01\textwidth}
               \includegraphics[width=0.42\textwidth,clip=]{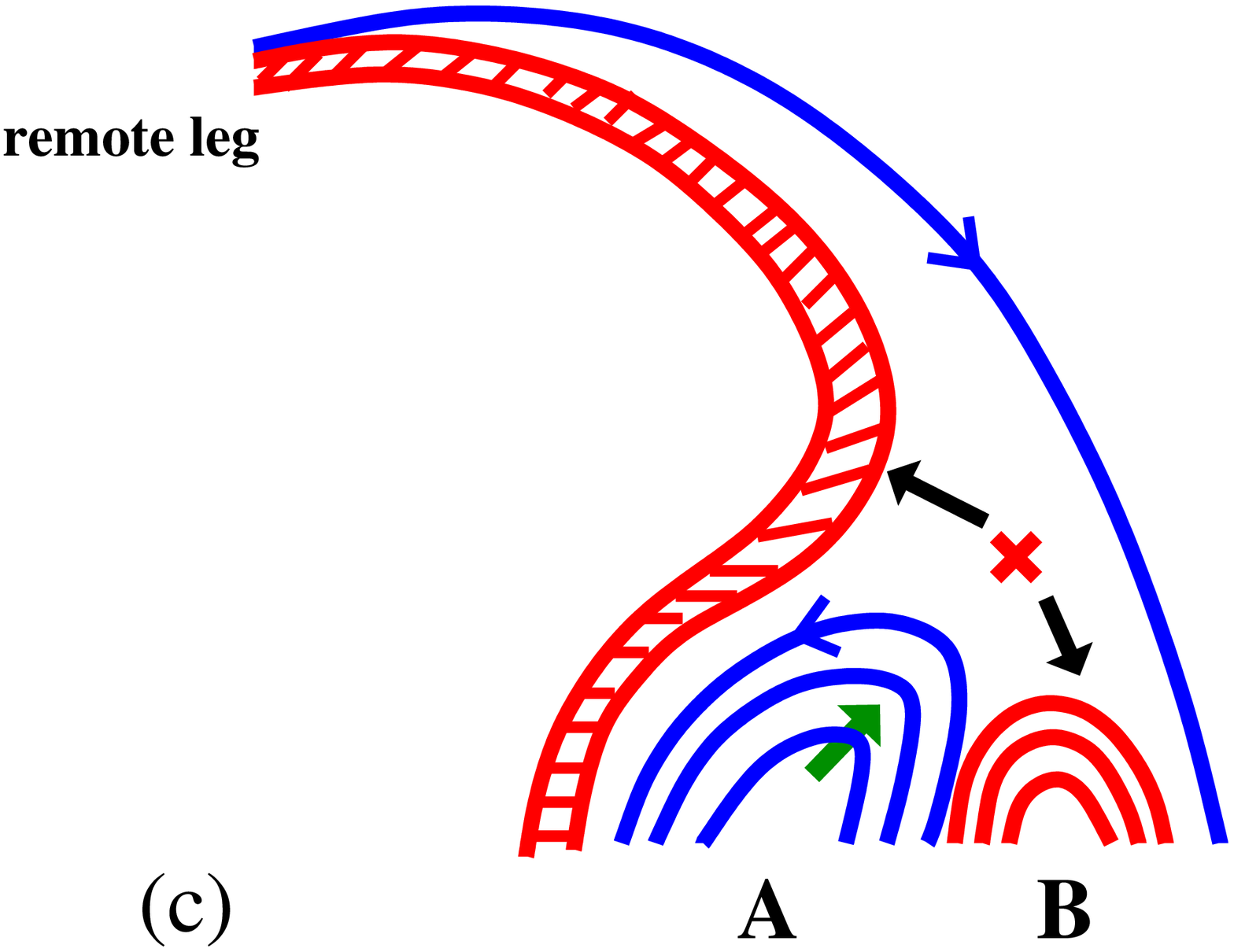}
               \hspace*{0.01\textwidth}
               \includegraphics[width=0.44\textwidth,clip=]{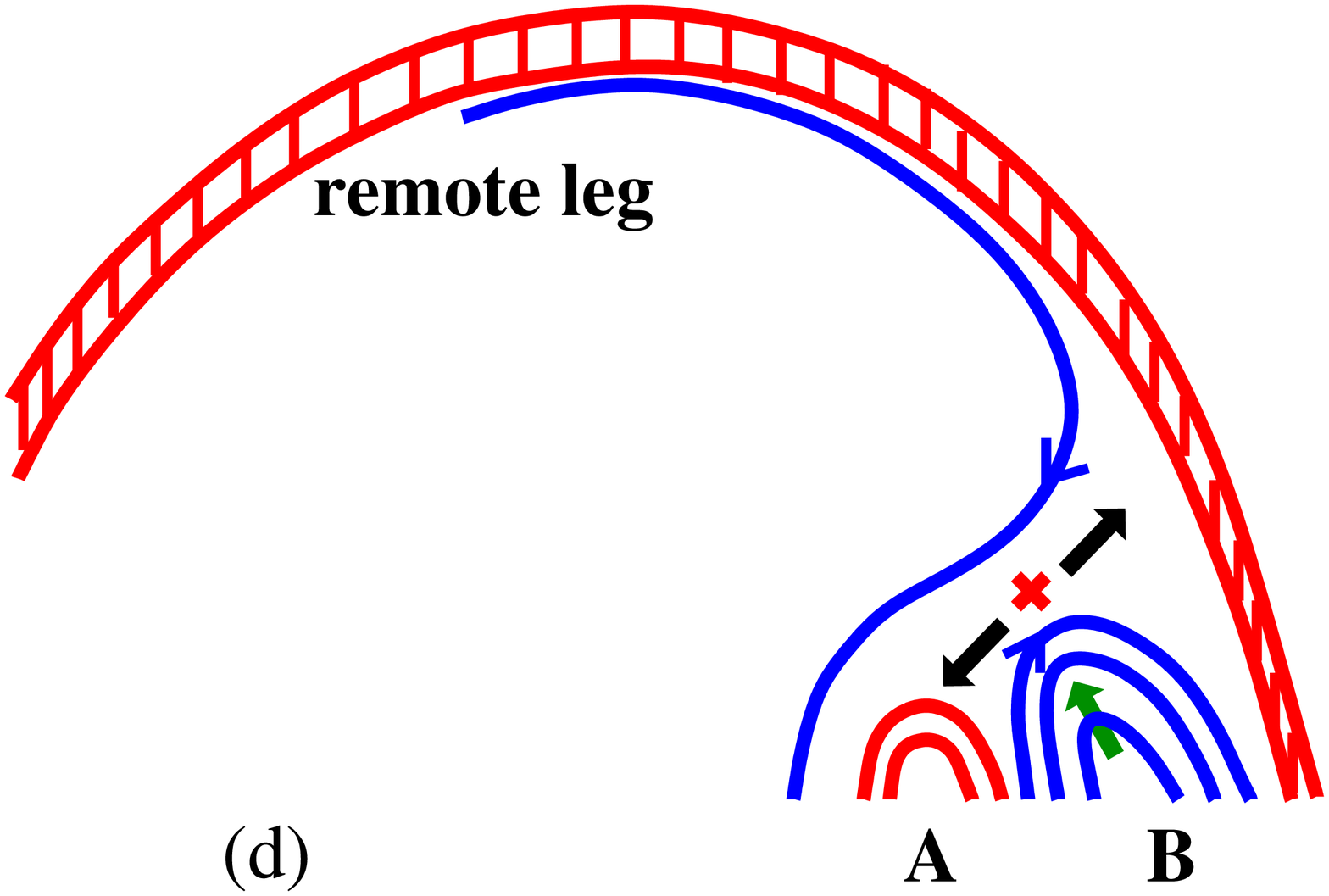}
              }
\caption{{\it Top panels}: SXR images of the CBP event observed at 23:08 UT 
on August 22 ({\it left}) and at 01:08 UT on August 23 ({\it right}). ``Arc 1'',
``Arc 2'', PA, PB, and CBP have the same meanings as in Fig.~\ref{fig5}.
{\it Bottom panels}: Cartoons that illustrate the weak ({\it left}) and strong ({\it
right}) flash phases of the CBP event. The blue/red solid lines represent
magnetic field lines in the inflow/outflow regions. The red crosses denote the
X-point of the reconnection region. The black arrows point to the direction of
mass motion. ``A'' and ``B'' stand for the magnetic field lines in the two 
patches of the CBP.}
\label{fig8}
\end{figure}

Based on the fact that the large-scale coronal loop drifted from the right to the left of
the CBP, the magnetic configuration responsible for the weak flash phase in 
Fig.~\ref{fig2} can be conceived and illustrated by
Fig.~\ref{fig8}c: As the small bipolar loop ``A'' expands upward, it
reconnects with the large-scale magnetic arc, producing bright loop ``B'' as
the reconnected loop below the X-point. At the same time, the large-scale coronal loop,
which is initially rooted on the right side of the two chambers as indicated by the
long blue line, is now rooted on the left side of the two chambers after reconnection
as indicated by the red thatched field lines. For comparison, one snapshot
of the observation is presented in Fig.~\ref{fig8}a. It is seen that all the
components in the reconnection model, i.e., loops A and B, and the thatched
arc, resemble the observation very well. Therefore, the expanding bipolar loop
in patch A corresponds to the reconnection inflow region during the weak flash
phase and patch B corresponds to the reconnected loop in the outflow region.

During the strong flash phase, patch B with a size of $\sim$15$\arcsec$
brightened first as it expanded. Based on the fact that the 
large-scale coronal loop drifted from left to the right of the CBP,
the corresponding magnetic configuration can be conceived and illustrated
by Fig.~\ref{fig8}d: As the bipolar loop B 
expands, it reconnects with the large-scale arc, producing bright patch A below 
the reconnection X-point. At the same time, the large-scale bright arc,
which is rooted on the left side of the two chambers as indicated by the blue
field line, is now rooted on the right side of the two chambers after reconnection
as indicated by the red thatched field lines. For comparison, one snapshot of 
the observation is shown in Fig.~\ref{fig8}b. Therefore, in the strong flash
phase, patch B corresponds to the expanding bipolar field in the reconnection
inflow region and patch A corresponds to the reconnected loop in the outflow
region, which is opposite to the case in the weak flash phase.

It was proposed that fast-mode waves could modulate the null-point 
reconnection in a way that one pair of anti-parallel magnetic field lines 
serve as reconnection inflow and outflow successively 
(Craig \& McClymont \cite{crai91}; McLaughlin et al. \cite{mcl09}), 
which was called ``oscillatory reconnection'' and reproduced in the 
magnetohydrodynamic (MHD) numerical simulations of interaction 
between emerging flux and pre-existing magnetic fields in coronal 
holes (Murray et al. \cite{mur09}; Archontis et al. \cite{arc10}). 
Enhanced gas pressure in the reconnection outflow region becomes 
higher than that in the inflow region, forcing magnetic field lines in the two 
bounded outflow regions to reconnect reversely. As a result, an oscillatory 
reconnection and recurrent outflows (jets) are formed. 
In this paper, we report such a process in a CBP for the first time. 
In order to distinguish this phenomenon from the wave-modulated 
reconnection where reconnection rate is modulated by waves but
the inflow and the outflow do not reverse (Aschwanden et al. \cite{asch94}; 
Chen \& Priest \cite{chen06}; Nakariakov et al. \cite{naka06}), we
call this phenomenon ``reciprocatory reconnection''. As illustrated by Fig.
\ref{fig8}, the successive brightenings of the CBP, with interval of $\sim$110 
min that is close to twice the time interval of the recurrent CBP flashes 
($\sim$120 min) studied by Zhang et al. (\cite{zqm12}), fit the reciprocatory
reconnection very well. The difference between the simulations of 
Murray et al. (\cite{mur09}) and our observation is that the reconnection in 
Murray et al. (\cite{mur09}) is a 
pure dissipative process where the reconnection rate decreases with time.
This is understandable since spontaneous magnetic reconnection is a
dissipative process, the magnetic fields after reconnection contains less
magnetic free energy. In our case, however, the highest reconnection rate
appeared in the final reconnection stage (01:00$-$01:10 UT on August 23),
implying that extra magnetic energy must be pumped into patch B after the 
weak flash phase. A possible mechanism of energy pumping is the twisting 
motion of the photosphere. After reconnection in the first flash phase, 
the reconnected field (patch B in Fig. \ref{fig8}c) is close to potential field.
However, during the following two hours, the everlasting convective motion 
in the solar photosphere may twist the bipolar loop in patch B, transferring
Poynting flux into the coronal part. Such a twisting motion drives the bipolar
loop in patch B to expand and then to reconnect with the overlying magnetic
field. The reconnected field lines become the newly-ignited patch A and the
thatched large loop in Fig.~\ref{fig8}d. The twisting motion-driven expansion 
and reconnection were numerically simulated by Pariat et al. (\cite{pari09}).

It has been revealed that a CBP is not just 
a simple bright point, it is often composed of a bundle of detailed sub-structures 
(Sheeley \& Golub \cite{shee79}; Dere \cite{dere08}). Then, a question is raised: 
What is the relation among the different parts of a CBP? Thanks to the high-resolution
observations from Hinode/XRT, we are able to discover the sub-structures of a CBP 
and their evolution. After careful analysis, we found that the CBP consisted of two 
small bright loops, patches A and B. They were ignited alternatively via reciprocatory 
reconnection. During the first flash phase, patch A corresponded to 
the reconnection inflow, whereas patch B to the reconnection outflow. However, 110 
min later, patch B became the reconnection inflow, whereas patch A became the 
reconnection outflow during the second flash phase. The transverse drifts of the
large-scale arc during the CBP flashes provide plausible evidence that the magnetic 
fields underwent rearrangement. Combining the time delays between the peaks of the 
two patches with the opposite directions of drifts of the large-scale arc, it is further 
concluded that reciprocatory reconnection took place in the CBP. Assuming that the
coronal Alfv\'en speed ($v_A$) is $\sim$1000 km s$^{-1}$, the reconnection
rate ($v_{in}/v_{A}$) in this CBP is roughly estimated to be around 0.01.

Generally speaking, when magnetic reconnection happens, say, in solar flares, 
the reconnection inflow region has the normal temperature, which is nearly the
same as the background corona, whereas the reconnection outflow region is
heated. However, for CBPs, the situation is different. Taking the first flash
phase for example, its reconnection process can be well illustrated by
Fig.~\ref{fig8}c, where patch A corresponds to the reconnection inflow region 
and patch B corresponds to the reconnection outflow region. However, Fig.
\ref{fig2} indicates that patch A also brightened, and its brightness reached
maximum even $\sim$2 min before that of patch B. Based on the MHD simulations
of Pariat et al. (\cite{pari09}) and the observational study of Zhang et al. (\cite{zqm12}), 
we conjecture that the brightening in the reconnection outflow (i.e., patch B) 
can be explained by the standard reconnection model for jets and CBPs 
(Shibata et al. \cite{shib92}), whereas the brightening in the reconnection inflow
(e.g., patch A) is due to QSL reconnection inside the twisting loop system itself 
as the convective  photosphere drags the loop system as simulated by Pariat
et al. (\cite{pari09}). Recently, Moreno-Insertis \& Galsgaard (\cite{more13}) performed
3D numerical simulations of jet eruptions as a result of magnetic reconnection between
an emerging flux rope and the pre-existing coronal magnetic field. In their numerical
results no brightening can be seen in the reconnection inflow region, i.e., the emerging
magnetic flux. The reason is probably that no twisting photospheric motion is exerted to
the inflow region, whereas rotating motion is introduced in the bottom boundary 
conditions in Pariat et al. (\cite{pari09}).

In an embedded bipolar magnetic configuration, a spine links the magnetic null point 
to infinity (Fletcher et al. \cite{fle01}; Pariat et al. \cite{pari10}; Liu et al. \cite{liu11}; 
Jiang et al. \cite{jiang13}) or to a remote place on the solar surface (Filippov \cite{fili99}; 
Masson et al. \cite{maso09}; Wang \& Liu \cite{wang12}; Sun et al. \cite{sun13}). 
In the latter case, the released thermal energy and accelerated particles during 
reconnection would be transferred along the magnetic spine to heat the 
chromosphere and produce brightening at the other footpoint of spine, which has 
been discovered in flares (Biesecker \& Thompson \cite{bies00}; Wang et al. \cite{wang01}; 
Moon et al. \cite{moon02}; Brosius \& Holman \cite{bro07}; Su et al. \cite{su13}). 
Using the high-cadence Hinode/XRT observations, similar phenomenon can be 
discovered in CBPs, as shown in Figs.~\ref{fig2} and \ref{fig5} and the online movies. 
The peak SXR intensities of the patch B in the CBP and the remote leg are nearly 
simultaneous. Considering the time cadence of the observation (32 s) and 
the distance between the remote brightening and the CBP ($\sim$35$\arcsec$), 
we can estimate the speed of energy transfer, which is $>$800 km s$^{-1}$. 
Hence, the energy transfer is due to either thermal conduction or non-thermal 
particles.

In this paper, we performed a case study of the CBP event observed by 
Hinode/XRT on 2009 August 22$-$23. The CBP
experienced two successive flashes, the first one was weak and the second one
stronger. During the flashes, the appearances of a two-chamber structure and 
a drifting large-scale arc convincingly support the classical magnetic 
reconnection model where an expanding loop reconnects with the pre-existing 
magnetic fields. The large-scale arc drifted leftwards (rightwards) during the 
weak (strong) flash, with apparent speeds of 16 km s$^{-1}$ and 4 
km s$^{-1}$, respectively. To the best of our knowledge, 
for the first time we discovered reciprocatory reconnection in a CBP event, 
i.e., reconnected loops in the outflow region of the first reconnection 
process serve as the inflow of the second reconnection process. The time delay 
between the peak SXR intensities of the outflow and inflow regions was $\sim$2 
min during each phase. We also proposed a scenario to interpret the physical 
relation between the sub-structures within a CBP. Additional case studies are 
required to testify the scenario using the high-cadence, high-resolution, and 
multi-wavelength observations from the Solar Dynamic Observatory (SDO).
MHD numerical simulations are also worthwhile to figure out
the mechanism of reciprocatory magnetic reconnection in CBPs.

\begin{acknowledgements}
The authors appreciate the referee for inspiring comments and suggestions.
Q.M.Z is grateful to Y. Guo and the solar physics group in Purple Mountain 
Observatory for valuable discussion. Hinode is a Japanese mission 
developed and launched by ISAS/JAXA, with NAOJ as domestic partner 
and NASA and STFC (UK) as international partners. This research is 
supported by the Chinese foundations NSFC (11303101, 11333009, 
11173062, 11025314, 11373023, and 11221063) and by 973 program 
2011CB811402.
\end{acknowledgements}

\end{document}